\documentclass{article}
\usepackage{graphicx}
\usepackage{float}
\usepackage{hyperref}
\DeclareUnicodeCharacter{200B}{ }
\usepackage[margin=1in]{geometry}
\usepackage{authblk}

\title{White paper: The Helix Pathogenicity Prediction Platform}
\author[1,2,3,*]{Bas Vroling}
\affil[1]{\small{These authors contributed equally towards this work}}
\affil[2]{Bio-Prodict, Nijmegen, the Netherlands}
\affil[3]{Centre for Molecular and Biomolecular Informatics (CMBI),\protect\\  Radboud University Medical Center, Nijmegen, the Netherlands}
\affil[*]{Correspondence: bvroling@bio-prodict.nl}
\author[1,2]{Stephan Heijl}
\date{}

\begin{document}
\maketitle

\abstract{
In this white paper we introduce Helix, an AI based solution for missense
pathogenicity prediction. With recent advances in the sequencing of
human genomes, massive amounts of genetic data have become available.
This has shifted the burden of labor for genetic diagnostics and
research from the gathering of data to its interpretation. Helix
presents a state of the art platform for pathogenicity prediction
of human missense variants. In addition to offering best-in-class
predictive performance, Helix offers a platform that allows researchers
to analyze and interpret variants in depth that can be accessed 
at \href{https://helixlabs.ai}{helixlabs.ai}.
}

\section{Introduction}\label{introduction}

Millions of human genomes and exomes have been sequenced, but their
clinical application remains limited due to the difficulty of
distinguishing rare disease-causing variants from benign genetic
variation. Without clear lines of evidence that show that a missense
variant is either benign or neutral, variants are deemed `variants of
uncertain significance', or VUS. The large number and rarity of these
VUSes presents a formidable obstacle, as the interpretation of these
variants is a difficult and time-consuming task, thereby severely
limiting the actionability of diagnostic sequencing. Furthermore, the
large amount of VUSes hinder the effective implementation of
applications that require large amounts of genotype-phenotype
associations such as disease association studies and patient
stratification in drug development programs.

To deal with this ever-growing problem, a large number of computational
tools have been developed that aim to predict the pathogenicity of novel
variants. These tools are typically based on limited data describing
evolutionary conservation and constraints, sometimes combined with the
physicochemical characteristics of the amino acids and information about
the structural domains in which these residues reside. The emergence of
many `meta-predictors' (or `ensemble predictors'; predictors that take
predictions of other predictors as inputs) has led to an increase in
overall predictive performance. However, despite obvious progress, these
tools leave room for improvement, especially in a clinical context.

At the core of missense variant prediction lies the problem of
understanding the role of proteins and their constituent amino acids. A
good understanding of this relation between sequence, structure and
function is the ultimate goal that enables insight in the impact of
missense variants.

Recently, advances have been made in closing this sequence-function gap
by artificial intelligence (AI) methods, where contextual language
models have been developed that can provide information about how
proteins are formed, shaped and function based on protein sequences
alone\cite{elnaggar_prottrans_2020}.
Trained on billions of protein sequences, these models contain a wealth
of information and can be applied to a wide range of predictive tasks
with just one protein sequence as input. However, the predictive power
of models solely based on these protein representations on explicit
protein prediction tasks still fall short of predictive methods that
include evolutionary information in the form of multiple sequence
alignments (MSAs) or information derived thereof.

This evolutionary information (e.g., amino acid conservation patterns)
can be used because functional and structural requirements lead to
specific selective pressures on proteins. In turn, identifying these
evolutionary constraints provides data and insights that can be used for
missense variant effect prediction and interpretation.

The ability to detect and interpret these evolutionary signals is
dependent on the quality of the data that is used. The application of
deep and high-quality multiple sequence alignments integrated with
protein structure data is commonplace in a number of fields that require
detailed understanding of the role of individual amino acids. Protein
optimization pipelines and drug development programs often critically
depend on the availability of deep annotations and integration of
protein sequence and structure data. Such high-quality protein
annotation data has not been available at scale, and the lack of such
data at large scale has hampered its application for missense effect
prediction.

We present Helix, a missense variant effect predictor built on a
resource of deep integrated protein data, in which over 30,000 protein
family structure-based MSAs, over 50,000 protein structures and
$\sim$60,000 human targeted sequence-based MSAs are combined
to provide deep annotations for all human proteins at the level of
individual amino acids. Helix incorporates this data in combination with
Contextual Language Model representations that contain implicit information about how
proteins are formed, shaped and function.

The features used by Helix describe proteins and their constraints at
multiple levels. This includes extensive MSA-derived metrics describing
selective pressures that act on proteins as a whole, on networks of
residues required for specific functions and on individual positions.
This information is integrated with features describing structural
aspects, e.g. secondary structure, flexibility and solvent-accessible
surfaces, both at the level of individual protein structures, as well as
in the form of aggregated data across protein domains and families.
Additionally, Helix uses descriptors of gene mutation tolerance, gene
occurrence throughout the evolutionary tree as well as data describing
protein-protein interactions. Best in class classification systems
trained on this data are combined with state of the art AI methodologies
to produce a predictor that offers the best of both worlds.

Helix was trained on a large set of well-annotated variants, using a
strict training regime where circularity is actively avoided. We
benchmark the performance of Helix using 10-fold cross validation datasets and
two previously published, clinically relevant datasets. We show that
Helix consistently outperforms existing predictors.

Predictions, however, are just that. For the effective use of these
predictions more context is often required. To this end, Helix provides high-quality 
predictions together with detailed variant reports that contain references to relevant scientific literature, 
information about evolutionary constraints, underlying data quality 
assessment and interactive protein structures.

\section{Methods}\label{methods}

Helix is built on top of the proprietary 3DM platform. 3DM is a protein
data and analytics platform that collects, combines and integrates
protein data for protein (super-)families. Examples of data included for
every protein family are protein sequences and structures, SNP data from
whole genome sequencing studies, scientific literature and patents, and
data derived thereof. Built around uniquely deep and high quality
structure based multiple sequence alignments, the 3DM platform offers
data and tooling that enable researchers to efficiently investigate the
relation between protein sequence, structure and function. Details about
the methodology behind 3DM can be found in Kuipers et al.\cite{kuipers_3dm_2010}.

Currently there are over 30.000 3DM protein family systems available.
These are built using templates selected from a) the SCOP
classifications of protein structures, and b) by clustering the
structures in the PDB. Overall, these cover the complete structural
space, and serve as an extremely rich data source with which to annotate
all human protein sequences. In short, this resource contains
high-quality MSA data integrated with protein structure data, scientific
literature, SNP data, etc for all parts of sequence space that can be
mapped to an existing protein structure. In terms of human proteins,
this resource can be used to annotate any part of a human protein that
can be aligned with a (not necessarily human) protein structure.

\subsection{Annotating the human
proteome}\label{annotating-the-human-proteome}

To enable predictions for the complete human proteome, it is necessary
that every individual amino acid is annotated. Since there are many
(parts of) proteins for which no structure information is available and
therefore cannot be annotated with the data present in the 3DM platform,
full-length sequence based alignments are a necessity to ensure full
sequence coverage. Therefore, Helix bases predictions not just on a
single alignment - where possible multiple alignments with different
depths are used to provide optimal predictive power. All positions in
the human exome are covered by full length, sequence-based multiple
sequence alignments (MSAs). In addition, (PFAM) domains are extracted
and covered by deeper, more specific alignments. In areas for which a
mapping to a protein structure can be made, the deep structure-based
alignments (and associated data) from the 3DM platform are used. In
typical sequence based MSAs there is a very distinct tradeoff between
coverage, alignment depth and alignment quality. In contrast, by using
multiple MSAs the requirement for full sequence coverage is fulfilled,
while maximising alignment quality and depth at the same time.

\begin{figure}[H]
\begin{center}
	\includegraphics[width=0.96\textwidth]{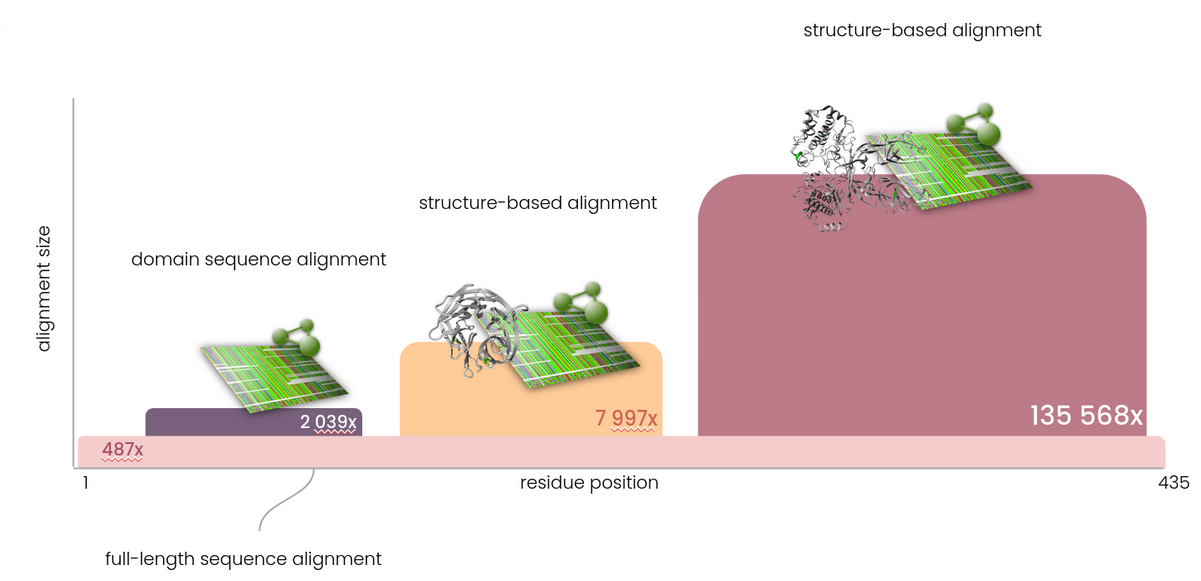}
\end{center}
	\caption{Illustration of the principles behind using
multiple alignments for annotating a single human protein. Every single
position is covered by the full length alignment, while select areas are
enhanced by using domain or structure-based alignments.}
\end{figure}

\subsubsection{Features}\label{features}

The features used in our tabular data models are hand curated and
include a large number of descriptors of evolutionary pressure that are
extracted from multiple sequence alignments. These include several
indicators of evolutionary pressure, acting on proteins as a whole, on
networks of residues required for specific functions and on individual
positions. Metrics include measures of conservation and change, entropy
metrics, correlated behaviour and where available, extensive structural
data (e.g. solvent exposure, electrostatic interactions, hydrogen
bonding) that describe the structural aspects of individual amino acids.
The feature set is expanded with a set of indicators of vulnerability on
the gene level. Minor allele frequency is explicitly not used as a
feature, as this would introduce a large amount of circularity and
predictive performance on rare variants would be very limited as a
result. Predictions from other classifiers are not used as a feature,
meaning that Helix is built from the ground up without suffering from
restrictions on the training data.

{
\subsubsection{Datasets and data preprocessing
}\label{datasets-and-data-preprocessing}}

The reference variant dataset used in this whitepaper was composed of
variants from ClinVar\cite{landrum_clinvar_2018}
and gnomAD\cite{genome_aggregation_database_consortium_mutational_2020}\@
and a dataset maintained by the Dutch genome diagnostic laboratories (VKGL)\cite{noauthor_vkgl_2019}.
ClinVar variants were included if the Clinvar review status was one star or higher, 
excluding variants with conflicting interpretations. ClinVar variants with 
``Benign'' and ``Likely benign'' interpretations were
included and labeled as benign variants, whereas those variants with
``Pathogenic'' and ``Likely pathogenic'' interpretations were included
and labeled as pathogenic variants. gnomAD variants were selected and
labeled benign if the minor allele frequency exceeded 0.1\%.

{
\subsection{Machine learning}\label{machine-learning}}

10-fold cross validation was used to train and analyze model
performance. In this process, the training data is divided into 10
partitions, where repeatedly 9 partitions are used for training and one
for validation. The details of how this partitioning of the data takes
place has important consequences for model performance and
generalizability to real-world applications.

Most practical implementations of the 10-fold cross validation technique
use a random split, where the selection of variants that end up in the
training or validation set for each fold is performed completely random,
irrespective of gene or residue positions. As we have shown in previous
research\cite{heijl_mind_2020}
this leads to issues with circularity and artificially increases
validation scores. In order to mitigate inflated scores, we use a gene
split strategy, where variants present in training and validation sets
are separated by gene; i.e. all variants for any given gene are either
present in the training set or in the validation set. Variants present
in the test sets
(BRCA1\cite{findlay_accurate_2018},
Clinical\cite{gunning_assessing_2020})
were excluded from the training/validation data and exclusively used for
testing.

\subsubsection{Ensemble prediction}

Using a variety of models to predict pathogenicity and combining their
predictions into an ensemble allows for better predictive performance
than using a single
classifier\cite{opitz_popular_1999}.
This process is used to combine all our models, while restricting the
training and test set combination to prohibit any contamination between
datasets.

Models used in the ensemble prediction include state of the art Gradient
Boosting classifiers, custom neural networks that perform well on
tabular data and proprietary contextual language models. The latter are
trained on a multitude of subtasks to infuse the model with external
data related to the pathogenicity prediction problem.

{
\section{Results}\label{results}}

The primary metric that was used for performance assessment is Matthew's
Correlation Coefficient. MCC is robust in the face of imbalanced
datasets and will report low scores for naive results. Comparisons are
made with a number of commonly used baseline, as well as newer
predictors. The former includes
SIFT\cite{noauthor_sift_2019},
CADD\cite{rentzsch_cadd_2019}
and
PolyPhen2\cite{adzhubei_method_2010},
while the latter consists of
REVEL\cite{ioannidis_revel_2016}
and
VEST4\cite{carter_identifying_2013}.

Figure 2 shows the Helix performance on the validation set. Helix
predictions were obtained by using 10-fold cross validation. Prediction
scores were obtained by testing on unseen genes, a more stringent way of
testing compared to the evaluation of the other tools listed. To
minimize training bias effects and ensure an as fair as possible
comparison, variants were excluded from this analysis if present in the
HGMD\cite{stenson_human_2003}
dataset or the PolyPhen HumDiv andHumVar training sets.

\begin{figure}[H]
\begin{center}
	\includegraphics[width=0.9\textwidth]{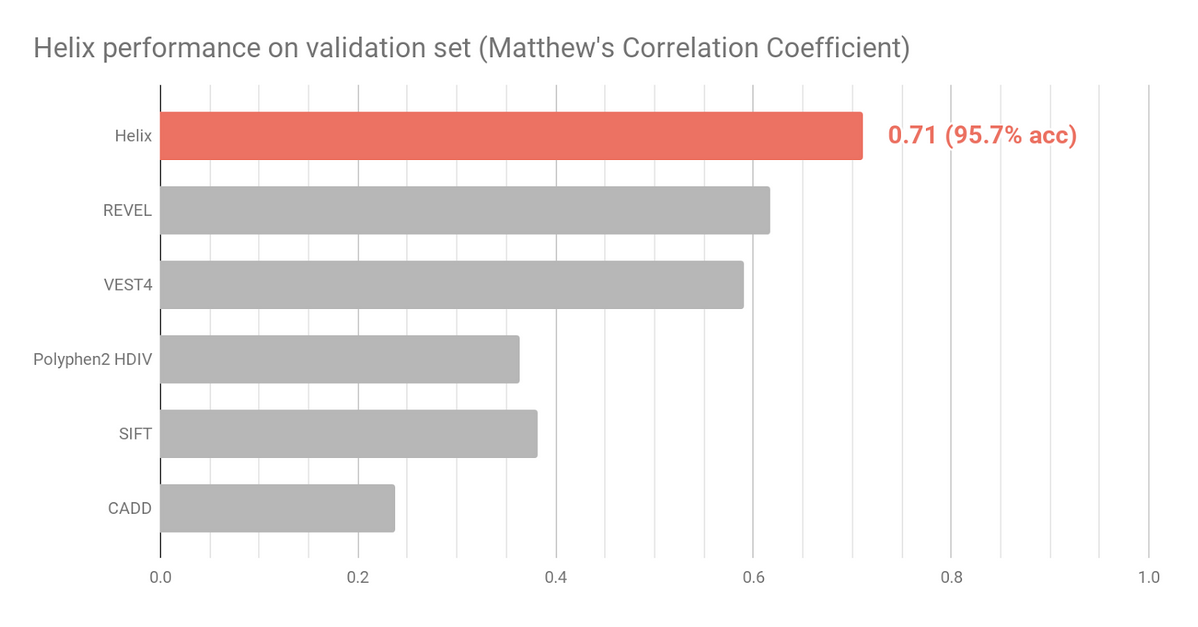}
\end{center}
	\caption{Helix performance on the 10-fold cross
validation set. This set covers over 400.000 variants in proteins the
model has not seen before.}
\end{figure}
{

Two datasets that more accurately reflect variants that might require
classification in a clinical setting were selected to independently
assess and compare Helix predictions (Figure 3 and Figure 4). Variants present in these sets
were excluded from the Helix training data. Overall, these datasets are
composed of variants that are clinically relevant, have been reviewed
recently, and where there is very limited overlap with training datasets
of existing predictive tools.

The Clinical dataset (n=1729) contains variants from the Deciphering Developmental Disorders
(DDD) study together with manually classified clinical variants and
variants identified in Amish individuals. Figure shows

\begin{figure}[H]
\begin{center}
	\includegraphics[width=0.9\textwidth]{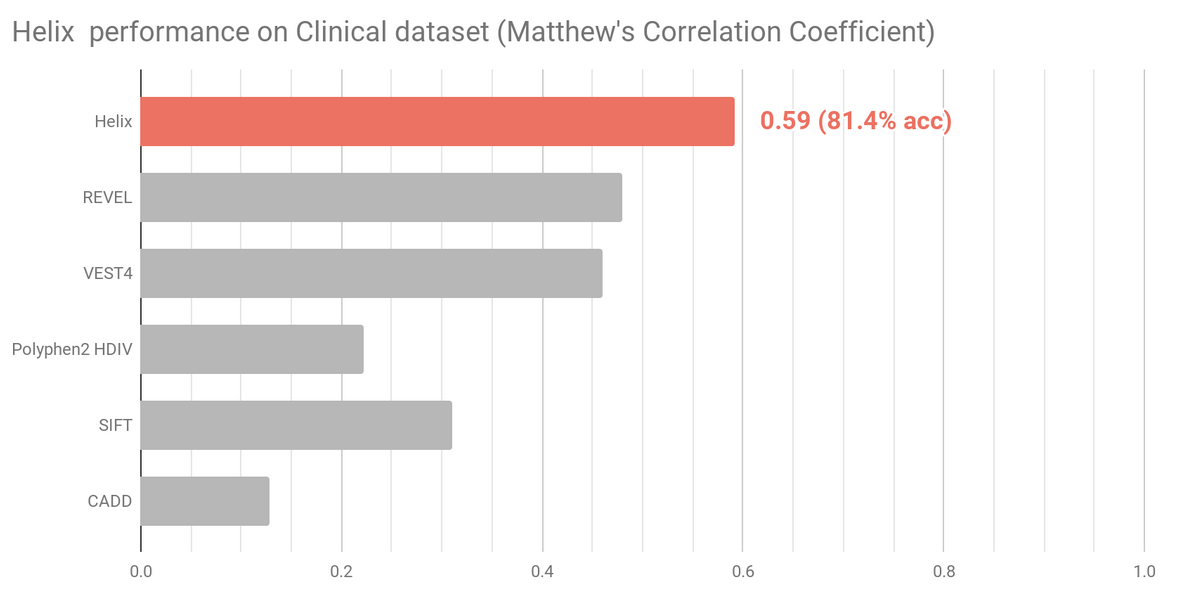}
\end{center}
	\caption{Helix performance on Clinical dataset.}
\end{figure}

The BRCA1 dataset (n=1605) contains experimentally assayed BRCA1 variants using
saturation genome editing.

\begin{figure}[H]
\begin{center}
	\includegraphics[width=0.9\textwidth]{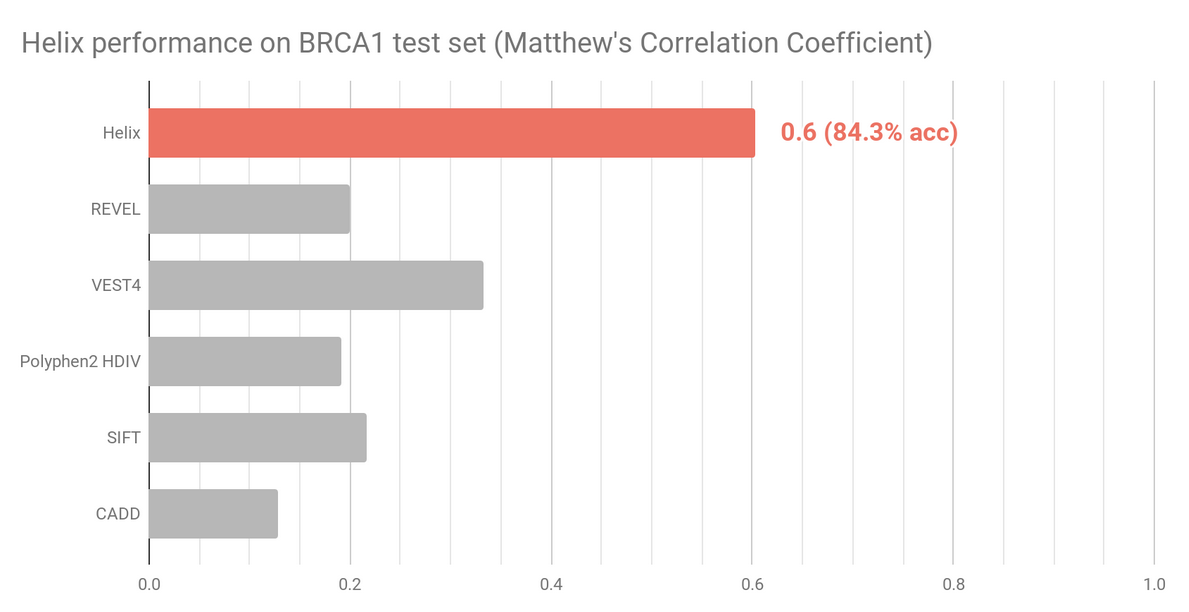}
\end{center}
	\caption{Helix performance on BRCA1 dataset.}
\end{figure}

{
\subsection{Reports}\label{reports}}

Helix predictions can be accessed through the web application, where
reports for every single variant in the human exome can be found.
Reports annotate variants with structural information, literature and
prediction data, with the ultimate goal of providing domain experts the
data and information they need to make the best decisions.

\begin{figure}[H]
\begin{center}
	\includegraphics[width=1.1\textwidth]{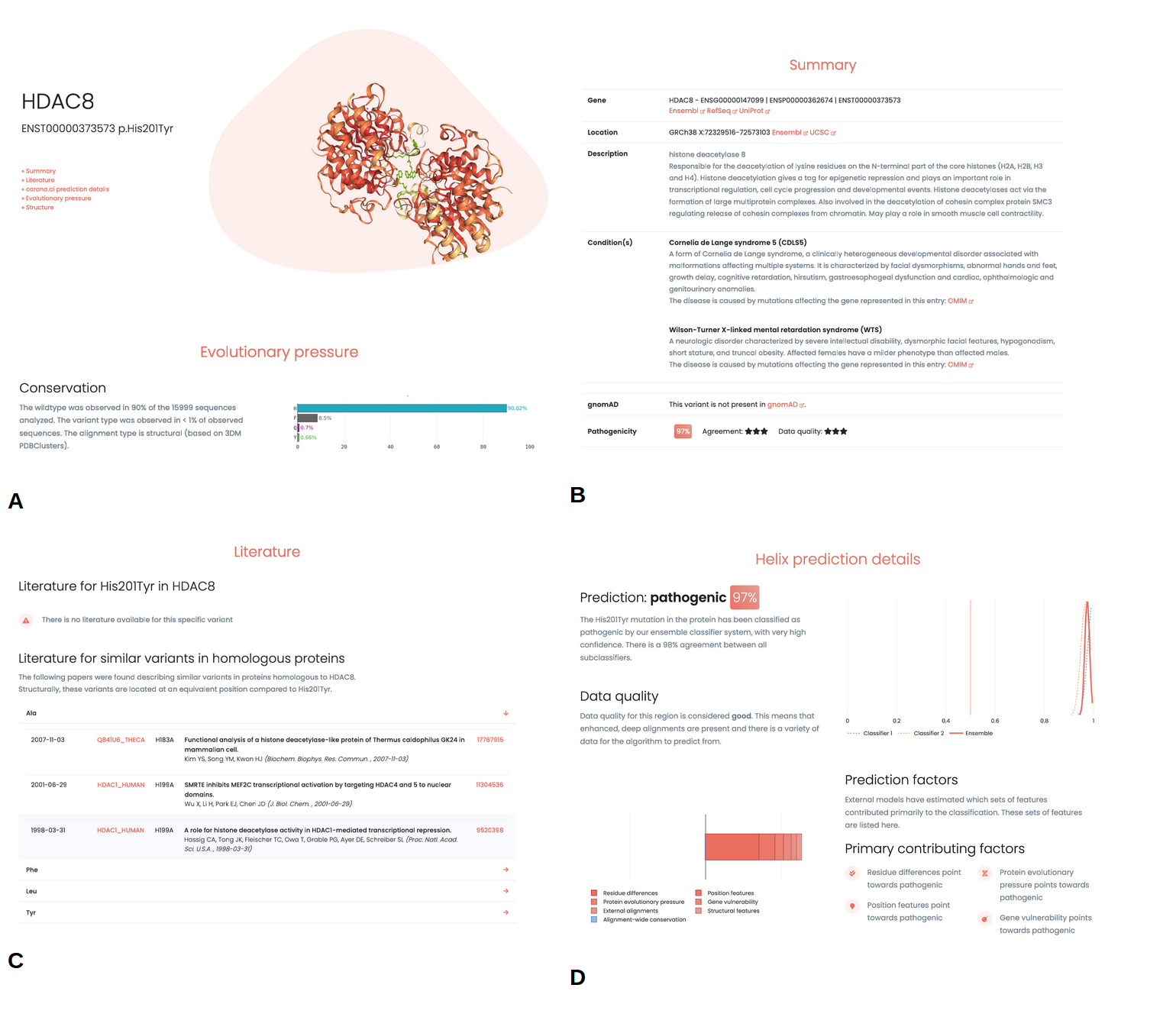}
\end{center}
	\caption{A sample report produced by Helix.}
\end{figure}

\hypertarget{variant-information}{%
\subsubsection{Variant information}\label{variant-information}}

WIthin the reports information is presented about the individual
variant. Known associated phenotypes for the gene are described,
together with relevant information about the affected gene and links to
external sources (e.g. ClinVar, gnomAD, UCSC, Ensembl). Furthermore,
information about the conservation for the wildtype and variant type
residue is presented, together with structural information (Figure 5 A, B).
Literature associated with the variant is listed. Importantly, this is
not limited to papers describing the exact variant but extends to
variants described in homologous proteins (Figure 5, C).

\hypertarget{prediction}{%
\subsubsection{Prediction}\label{prediction}}

The variant report provides details on various aspects of the
predictions (Figure 5, D). Agreement between the different sub-predictors
in the ensemble is shown, where a high agreement generally indicates
more certainty in the prediction.

A diverse set of metrics is used to present an intuitive score
indicating the quality of the data underlying the prediction. This score
is an indication of the amount of confidence one should place on a given
prediction. No matter how good the predictive model is, if the data that
is provided to the model is of limited quality (i.e. small MSAs, no
protein structures available), one should be careful in placing too much
confidence in these predictions.

In addition, the report offers a visual indication of the factors that
are estimated to contribute most to the individual predictions. Features
are grouped into a few categories to simplify interpretation.

\hypertarget{discussion}{%
\section{Discussion}\label{discussion}}

Helix is built on a unique resource of integrated protein data, and
represents a state-of-the-art predictive tool that outperforms currently
available predictors by a large margin. However, Helix is by no means
`finished'; many challenges remain open for the problem of predicting
the effects of missense variants. In fact, despite offering impressive
predictive performance, Helix is just getting started. Novel challenges
lie in problem domains such as accurate modeling of protein-protein
interactions and the integration of pathway information to be able to
assess downstream effects of variants within the context of larger
systems.

Helix offers an interactive platform that allows researchers to
investigate both the variants as the predictions in great detail. The
variant reports, in which variants and their associated predictions are
presented in a coherent fashion to add context and interpretability,
offer an important first step to help researchers make sense of all that
data.

In the future, Helix will continue to integrate novel technologies and
performance increasing features. Helix is supported by a longstanding
bioinformatics company that is dedicated to its further development. As
the AI field continues to provide new solutions that touch the variant
effect prediction space, we envision including these technologies with
even better performance and interpretability as a result.

\hypertarget{acknowledgments}{%
\section{Acknowledgments}\label{Acknowledgments}}
The research leading to these results has received funding from the European Union’s Horizon 2020 research and innovation programme under grant agreement No 634935 (BRIDGES), No 635595 (CarbaZymes) and No 685778 (VirusX).

\bibliographystyle{unsrt}
\bibliography{library}

\begin{thebibliography}{10}

\bibitem{elnaggar_prottrans_2020}
Ahmed Elnaggar, Michael Heinzinger, Christian Dallago, Ghalia Rehawi, Yu~Wang,
  Llion Jones, Tom Gibbs, Tamas Feher, Christoph Angerer, Martin Steinegger,
  Debsindhu Bhowmik, and Burkhard Rost.
\newblock {ProtTrans}: {Towards} {Cracking} the {Language} of {Life}’s {Code}
  {Through} {Self}-{Supervised} {Deep} {Learning} and {High} {Performance}
  {Computing}.
\newblock preprint, Bioinformatics, July 2020.

\bibitem{kuipers_3dm_2010}
Remko~K. Kuipers, Henk-Jan Joosten, Willem J.~H. van Berkel, Nicole G.~H.
  Leferink, Erik Rooijen, Erik Ittmann, Frank van Zimmeren, Helge Jochens, Uwe
  Bornscheuer, Gert Vriend, Vitor A. P.~Martins dos Santos, and Peter~J.
  Schaap.
\newblock {3DM}: systematic analysis of heterogeneous superfamily data to
  discover protein functionalities.
\newblock {\em Proteins}, 78(9):2101--2113, July 2010.

\bibitem{landrum_clinvar_2018}
Melissa~J. Landrum, Jennifer~M. Lee, Mark Benson, Garth~R. Brown, Chen Chao,
  Shanmuga Chitipiralla, Baoshan Gu, Jennifer Hart, Douglas Hoffman, Wonhee
  Jang, Karen Karapetyan, Kenneth Katz, Chunlei Liu, Zenith Maddipatla, Adriana
  Malheiro, Kurt McDaniel, Michael Ovetsky, George Riley, George Zhou,
  J.~Bradley Holmes, Brandi~L. Kattman, and Donna~R. Maglott.
\newblock {ClinVar}: improving access to variant interpretations and supporting
  evidence.
\newblock {\em Nucleic Acids Research}, 46(D1):D1062--D1067, January 2018.

\bibitem{genome_aggregation_database_consortium_mutational_2020}
{Genome Aggregation Database Consortium}, Konrad~J. Karczewski, Laurent~C.
  Francioli, Grace Tiao, Beryl~B. Cummings, Jessica Alföldi, Qingbo Wang,
  Ryan~L. Collins, Kristen~M. Laricchia, Andrea Ganna, Daniel~P. Birnbaum,
  Laura~D. Gauthier, Harrison Brand, Matthew Solomonson, Nicholas~A. Watts,
  Daniel Rhodes, Moriel Singer-Berk, Eleina~M. England, Eleanor~G. Seaby,
  Jack~A. Kosmicki, Raymond~K. Walters, Katherine Tashman, Yossi Farjoun, Eric
  Banks, Timothy Poterba, Arcturus Wang, Cotton Seed, Nicola Whiffin,
  Jessica~X. Chong, Kaitlin~E. Samocha, Emma Pierce-Hoffman, Zachary Zappala,
  Anne~H. O’Donnell-Luria, Eric~Vallabh Minikel, Ben Weisburd, Monkol Lek,
  James~S. Ware, Christopher Vittal, Irina~M. Armean, Louis Bergelson, Kristian
  Cibulskis, Kristen~M. Connolly, Miguel Covarrubias, Stacey Donnelly, Steven
  Ferriera, Stacey Gabriel, Jeff Gentry, Namrata Gupta, Thibault Jeandet, Diane
  Kaplan, Christopher Llanwarne, Ruchi Munshi, Sam Novod, Nikelle Petrillo,
  David Roazen, Valentin Ruano-Rubio, Andrea Saltzman, Molly Schleicher, Jose
  Soto, Kathleen Tibbetts, Charlotte Tolonen, Gordon Wade, Michael~E.
  Talkowski, Benjamin~M. Neale, Mark~J. Daly, and Daniel~G. MacArthur.
\newblock The mutational constraint spectrum quantified from variation in
  141,456 humans.
\newblock {\em Nature}, 581(7809):434--443, May 2020.

\bibitem{noauthor_vkgl_2019}
{VKGL} - {Vereniging} {Klinisch} {Genetische} {Laboratoriumdiagnostiek} -
  {Home}, March 2019.

\bibitem{heijl_mind_2020}
Stephan Heijl, Bas Vroling, Tom van~den Bergh, and Henk-Jan Joosten.
\newblock Mind the gap: preventing circularity in missense variant prediction.
\newblock preprint, Bioinformatics, May 2020.

\bibitem{findlay_accurate_2018}
Gregory~M. Findlay, Riza~M. Daza, Beth Martin, Melissa~D. Zhang, Anh~P. Leith,
  Molly Gasperini, Joseph~D. Janizek, Xingfan Huang, Lea~M. Starita, and Jay
  Shendure.
\newblock Accurate classification of {BRCA1} variants with saturation genome
  editing.
\newblock {\em Nature}, 562(7726):217--222, October 2018.

\bibitem{gunning_assessing_2020}
Adam~C Gunning, Verity Fryer, James Fasham, Andrew~H Crosby, Sian Ellard,
  Emma~L Baple, and Caroline~F Wright.
\newblock Assessing performance of pathogenicity predictors using clinically
  relevant variant datasets.
\newblock {\em Journal of Medical Genetics}, pages jmedgenet--2020--107003,
  August 2020.

\bibitem{opitz_popular_1999}
D.~Opitz and R.~Maclin.
\newblock Popular {Ensemble} {Methods}: {An} {Empirical} {Study}.
\newblock {\em Journal of Artificial Intelligence Research}, 11:169--198,
  August 1999.

\bibitem{noauthor_sift_2019}
{SIFT}: {Predicting} amino acid changes that affect protein function. -
  {PubMed} - {NCBI}, March 2019.

\bibitem{rentzsch_cadd_2019}
Philipp Rentzsch, Daniela Witten, Gregory~M Cooper, Jay Shendure, and Martin
  Kircher.
\newblock {CADD}: predicting the deleteriousness of variants throughout the
  human genome.
\newblock {\em Nucleic Acids Research}, 47(D1):D886--D894, January 2019.

\bibitem{adzhubei_method_2010}
Ivan~A. Adzhubei, Steffen Schmidt, Leonid Peshkin, Vasily~E. Ramensky, Anna
  Gerasimova, Peer Bork, Alexey~S. Kondrashov, and Shamil~R. Sunyaev.
\newblock A method and server for predicting damaging missense mutations.
\newblock {\em Nature Methods}, 7(4):248--249, April 2010.

\bibitem{ioannidis_revel_2016}
Nilah M. Ioannidis, Joseph H. Rothstein, Vikas Pejaver, Sumit Middha,
  Shannon K. McDonnell, Saurabh Baheti, Anthony Musolf, Qing Li, Emily
  Holzinger, Danielle Karyadi, Lisa A. Cannon-Albright, Craig C. Teerlink,
  Janet L. Stanford, William B. Isaacs, Jianfeng Xu, Kathleen A. Cooney,
  Ethan M. Lange, Johanna Schleutker, John D. Carpten, Isaac J. Powell,
  Olivier Cussenot, Geraldine Cancel-Tassin, Graham G. Giles, Robert J.
  MacInnis, Christiane Maier, Chih-Lin Hsieh, Fredrik Wiklund, William J.
  Catalona, William D. Foulkes, Diptasri Mandal, Rosalind A. Eeles, Zsofia
  Kote-Jarai, Carlos D. Bustamante, Daniel J. Schaid, Trevor Hastie,
  Elaine A. Ostrander, Joan E. Bailey-Wilson, Predrag Radivojac, Stephen N.
  Thibodeau, Alice S. Whittemore, and Weiva Sieh.
\newblock {REVEL}: {An} {Ensemble} {Method} for {Predicting} the
  {Pathogenicity} of {Rare} {Missense} {Variants}.
\newblock {\em American Journal of Human Genetics}, 99(4):877--885, October
  2016.
\newblock Number: 4.

\bibitem{carter_identifying_2013}
Hannah Carter, Christopher Douville, Peter~D Stenson, David~N Cooper, and
  Rachel Karchin.
\newblock Identifying {Mendelian} disease genes with the {Variant} {Effect}
  {Scoring} {Tool}.
\newblock {\em BMC Genomics}, 14(Suppl 3):S3, May 2013.
\newblock Number: Suppl 3.

\bibitem{stenson_human_2003}
Peter~D. Stenson, Edward~V. Ball, Matthew Mort, Andrew~D. Phillips,
  Jacqueline~A. Shiel, Nick~S.T. Thomas, Shaun Abeysinghe, Michael Krawczak,
  and David~N. Cooper.
\newblock Human {Gene} {Mutation} {Database} ({HGMD} $^{\textrm{®}}$ ): 2003
  update: {HGMD} 2003 {UPDATE}.
\newblock {\em Human Mutation}, 21(6):577--581, June 2003.

\end{thebibliography}

\end{document}